\documentclass[]{imsart}

\RequirePackage{amsthm,amsmath,amsfonts,amssymb}
\RequirePackage[authoryear]{natbib}
\RequirePackage[colorlinks,citecolor=blue,urlcolor=blue]{hyperref}
\RequirePackage{graphicx}
\usepackage{mathrsfs}  
\usepackage{tikz}
\usepackage{geometry}
\usepackage{tabularx}
\usetikzlibrary{matrix}

\tikzset
{
	over/.style={preaction={draw=white,-,line width=5pt}},
}

\startlocaldefs
\newcommand{\de}{\mathrm{d}}
\newcommand{\opt}[1]{\ensuremath{\left(#1\right)}}
\DeclareMathOperator{\Cov}{Cov}

\endlocaldefs

\begin{document}
	\graphicspath{{figs/}{figs/}}
	
	\begin{frontmatter}
		\title{An adaptive kernel estimator for the intensity function of spatio-temporal point processes}
		\runtitle{Intensity adaptive estimator for spatio-temporal point processes}
		
		\begin{aug}
			\author[A]{\fnms{Jonatan A.}~\snm{Gonz\'alez}\ead[label=e1]{jonathan.gonzalez@kaust.edu.sa}},
			\and
			\author[A]{\fnms{Paula}~\snm{Moraga}\ead[label=e2]{paula.moraga@kaust.edu.sa}}
			\address[A]{Computer, Electrical and Mathematical Sciences and Engineering Division, King Abdullah University of Science
				and Technology (KAUST), Thuwal 23955-6900, Saudi Arabia \printead[presep={,\ }]{e1,e2}}
			
		\end{aug}
		
		\begin{abstract}
			In spatio-temporal point pattern analysis, one of the main statistical objectives is to estimate the first-order intensity function, i.e., the expected number of points per unit area and unit time. This estimation is usually carried out non-parametrically through kernel functions, where one of the most frequent handicaps is the selection of kernel bandwidths prior to estimation. This work presents an intensity estimation mechanism in which the spatial and temporal bandwidths change at each data point in a spatio-temporal point pattern. This class of estimators is called adaptive estimators, and although there have been studied in spatial settings, little has been said about them in the spatio-temporal context. We define the adaptive intensity estimator in the spatio-temporal context and extend a partitioning technique based on the bandwidths quantiles to perform a fast estimation. We demonstrate through simulation that this technique works well in practice with the partition estimator approximating the direct estimator and much faster computation time. Finally, we apply our method to estimate the spatio-temporal intensity of fires in the Amazonia basin.
		\end{abstract}
		
		\begin{keyword}
			\kwd{Amazonia fires}
			\kwd{Bandwidth selection}
			\kwd{Intensity function}
			\kwd{Partitioning algorithm}
			\kwd{Spatio-temporal point process}
		\end{keyword}
		
	\end{frontmatter}
	
	\section{Introduction}\label{introduction}
	
	In spatio-temporal point processes, one of the essential characteristics of a given observation, that is, of a point pattern, is the first-order intensity, which corresponds to the localised expected number of points in the observation window and temporal interval \citep{gonzalezetal2016}. Knowing the intensity of a point process is often extremely important for interpretation since it represents the first moment, an analogy to the average of a population \citep{baddeley2015spatialR}, and therefore it is a fundamental descriptive characteristic. Additionally, this first-order statistic is a key ingredient in the subsequent estimation of some higher-order descriptors, such as those that measure the interaction between pairs of points \citep{gonzalezetal2016}. Furthermore, the intensity of one process can be related to the intensity of another process, obtaining a quantity commonly known as relative risk, which is used to investigate the fluctuations of one process compared to another \citep{davies2018relativerisksparr}.
	
	Kernel smoothing is a non-parametric technique classically used to estimate some types of functions, such as intensities and probability density functions. This technique is increasingly common for intensity estimation in spatial and spatio-temporal statistics, especially in those cases where additional information is unavailable to understand the distribution of points in space or space-time \citep{fernandohazelton2014risk}. 
	One of the main disadvantages of Kernel estimation is the need of prior knowledge of the bandwidth or smoothing parameter of the kernel used for estimation. Larger bandwidths provide smoother estimates and vice versa. Regardless of the data dimensionality, this smoothing parameter is fundamental for the adequate estimation of the intensity, and a wrong choice may have unfortunate consequences \citep[see e.g.\ ][and references therein]{baddeley2015spatialR}.
	In the statistical literature, it is widespread to use fixed bandwidth kernels, mainly due to the simplicity of the implementation and the steps involved. However, this classical approach's lack of spatial and temporal adaptability often results in poor estimation, especially in highly heterogeneous point processes with very complicated underlying features that affect the intensity structure within the study region \citep{Davies2018kernel,davies2018relativerisksparr}. The fixed kernel density estimator struggles to capture important finer details in crowded areas when much smoothing is applied to control noise where data is sparse. Conversely, if less smoothing is applied (to retain more of this detail), we can expect to see spurious bumps generated by isolated points in low-density regions.
	
\begin{figure}[h!tb]
		\centering
		\includegraphics[width=0.44 \linewidth]{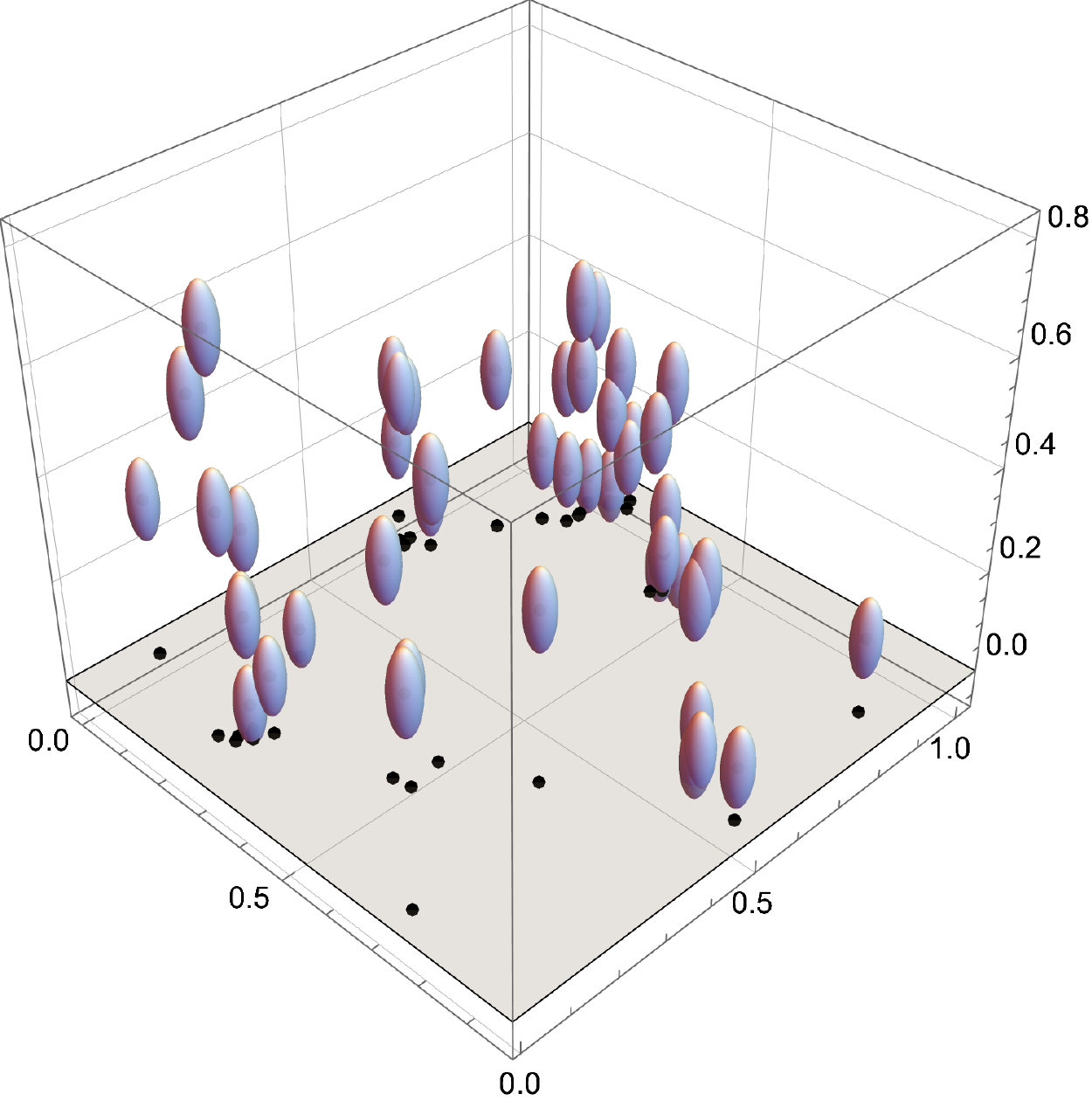} \hspace{13mm}
		\includegraphics[width=0.44 \linewidth]{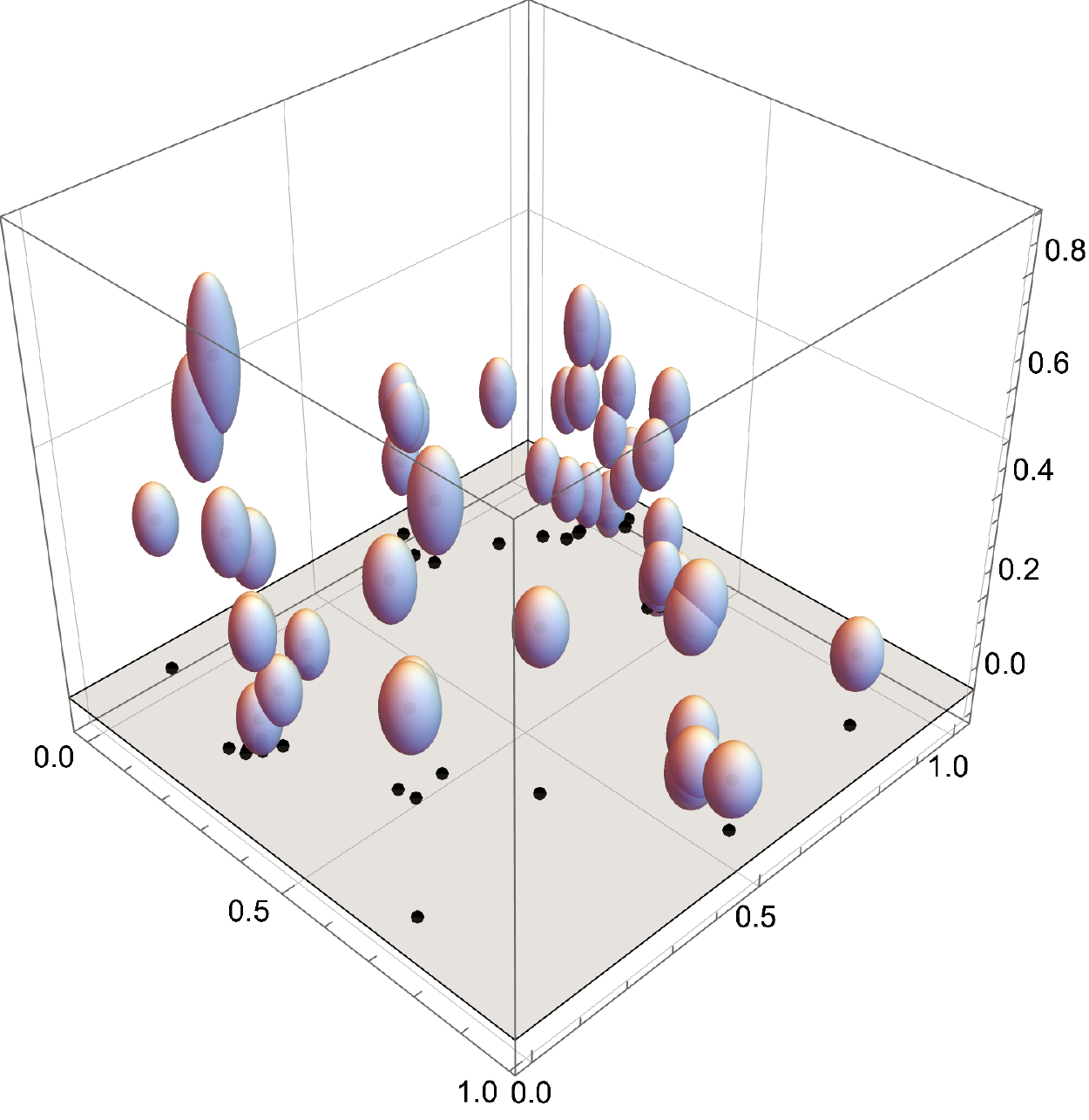}
		\caption{(Right) Spatio-temporal point pattern with fixed kernel bandwidth in space and time for each data point (drawn at one particular level as translucent isosurfaces). (Left) Spatio-temporal point pattern with adaptive kernel bandwidth spreading in space and time. In the latter image, the spatial projection of the points is shown where an increase in time is interpreted as moving up along the vertical axis.}
		\label{fig:ellipsoids}
	\end{figure}

	Therefore, a more intuitive approach emerges consisting of variable smoothing; in this technique, the amount of smoothing is inversely related to the density of the points. Known as\textit{ adaptive}, this smoothing has shown better bias levels \citep{abramson1982bandwidth,hallmarron1988variablewindow} and lower integrated square error \citep{diggleetal2005}. There are two different ways to approach variable bandwidth. The fitting can be performed at each location on the function evaluation grid, i.e., the spatio-temporal grid in which the intensity function is numerically evaluated. This technique is related to the so-called \textit{locally adaptive }or \textit{balloon} estimators. Alternatively, the rescaled bandwidth of the kernels can be associated with the data points; this technique is called the \textit{point fitting, sample smoothing}, or \textit{sample point }method \citep[see e.g.\ ][and references therein]{scott2015multivariate}. Following \cite{Davies2018kernel}, we focus on sample point estimators with the \textit{square root} methodology; this methodology assigns bandwidths proportional to the square root of the pilot estimates \citep{abramson1982bandwidth}. To illustrate the idea of point adaptation, Figure \ref{fig:ellipsoids} provides an example of a spatio-temporal point pattern. The ellipsoids correspond to a variable bandwidth in space and time for each point.
	We focus on the practical aspects of the methodology, combining theory with implementation. We provide the code to apply the methods discussed here in the newly developed software package \texttt{kernstadapt} (kernel adaptive estimation in the spatio-temporal context). The package is freely available for the \texttt{R} language on \texttt{GitHub} at \url{https://github.com/jagm03/kernstadapt}.
	
	This article is organised as follows. Section \ref{sec:basicpp} presents some basic concepts related to spatio-temporal point processes, the first-order intensity function, and its classical estimation through kernels. Section \ref{sec:stadaptive} explicitly defines the expression for the space-time intensity estimator in its adaptive version with two types of bandwidth, spatial and temporal; and the partition algorithm presented for fast estimation. Section \ref{sec:simulation} shows how our method works in terms of squared integrated errors and speed through simulation. In Section \ref{sec:amazonia}, we apply our method to a database of fire reports on Amazonia. Finally, we conclude with specific comments and possible future lines of research in Section \ref{sec:discussion}.
	
	\section{Spatio-temporal point processes in a nutshell}\label{sec:basicpp}
	We say that $X$ is a \textit{spatio-temporal point process} whenever it is a random countable subset of $\mathbb{R}^2 \times \mathbb{R}$, for which $|X \cap(A\times B)|<\infty$ for bounded $A\times B \subseteq \mathbb{R}^2 \times \mathbb{R}$; we use $|\cdot|$ to denote both set cardinality and absolute value of a real number. On the other hand, a spatio-temporal point pattern observed in $W\times T$ is a realisation of a spatio-temporal point process in $\mathbb{R}^2\times \mathbb{R}$ to $W\times T$. We denote $N(A\times B)$ as the number of points of a set $(A\times B )\cap X$, where $A \subseteq W$ and $B \subseteq T$. 
	
	We say that a spatio-temporal point process $X$ is  {\em (spatio-temporally) stationary} when shifted processes $(\mathbf{u},v)+X$ have the same distribution as the original process $X$ for any $(\mathbf{u},v)\in W\times T$.  $X$ is defined as {\em (spatially) isotropic} if, for any rotation ${\bf r}$ around the origin, every rotated point process ${\bf r} X=\{({\bf r}\mathbf{u},v) : (\mathbf{u},v)\in X\}$ has the same distribution as $X$ \citep[see e.g.\ ][]{Moller2012,gonzalezetal2016}.
	
	\subsection{Intensity}\label{subsectionintensityfunctions}
	The expected value of the number of points governs the univariate distributions of the points of $X$ in $W\times T$. A statistical analysis of spatio-temporal point patterns usually begins estimating and modelling the \textit{intensity function}, i.e., the expected number of points per unit area per unit time. When the intensity function $\lambda(\cdot)$ exists, the following identity holds,
	\begin{equation*}
		\mathbb{E}[N(A\times B)]= \int_{A \times B} \lambda(\mathbf{u},v) \de \mathbf{u} \de v.
	\end{equation*}
	When $X$ is stationary, also referred to as $X$ being \textit{homogeneous} (or completely stationary, see \citealp{illian2008}), then $\lambda(\mathbf{u},v)\equiv\lambda>0$. This constant is referred to as the {\em intensity} of $X$.
	
	\subsubsection{Separability}
	The separability of the intensity function of a spatio-temporal point process is a pragmatic and common assumption in the spatio-temporal point process literature \citep[see, e.g.,\ ][]{diggle2009,Moller2012,gabriel2013stpp,gonzalezetal2016}. 
	When the intensity function of a spatio-temporal temporal point process can be factorised (almost everywhere) as 
	\begin{equation*}\label{firstorderseparability}
		\lambda(\mathbf{u},v) = \lambda_1(\mathbf{u}) \lambda_2(v),
	\end{equation*}
	where 
	$\lambda_1(\cdot)$ and $\lambda_2(\cdot)$ are non-negative functions, the process is referred to as {\em first-order spatio-temporal separable}. Note that these functions are not unique. The separability assumption simplifies the analysis and the estimation; however, it rarely holds in practice, so it becomes essential to perform the estimation through non-separable estimators. Several works have focused on statistical tests to verify the separability assumption; see, e.g.,\ \cite{schoenberg2004,diazmateu2013,fuentes2018separabilityratio}, and more recently, \cite{ghorbani2021separability}.
	
	\subsubsection{Kernel estimator with fixed bandwidths}
	For non-parametric estimation of the spatio-temporal intensity function, not necessarily separable, it is common to use a kernel estimator  \citep[see, e.g.,\ ][and references therein]{fernandohazelton2014risk,gonzalezetal2016},
	\begin{equation}
		\hat{\lambda}\left(\mathbf{u},v\right) =\frac{1}{e \left(\mathbf{u},v\right) } \sum_{i=1}^{n}K^{\text{s}}_{\epsilon}\left(\mathbf{u}-\mathbf{u}_{i}\right) K^{\text{t}}_{\delta}\left(v-v_{i}\right),
		\qquad (\mathbf{u},v) \in W\times T,
		\label{eq:stkernelintensity}
	\end{equation}
	where $K^{\text{s}}_{\epsilon}(\cdot)$ and $K^{\text{t}}_{\delta}(\cdot)$ are bivariate and univariate kernels for space and time, $\epsilon, \delta >0$ are the \textit{bandwidths}, i.e., the smoothing parameters, and $e\left(\mathbf{u},v \right)$ is an edge-correction factor included in the estimation. The edge correction term intends to correct the bias that comes from the estimation at the edges \citep[see e.g.\ ][]{diggle1985kernel,Jones1993,} or preserve the estimator's mass, i.e., to guarantee that $\int \hat{\lambda} =n$  \citep{vanLieshout2011,ghorbani2013}. The uniform edge-correction factor can be defined as
	\begin{equation}\label{eq:stedgecorrection}
		e\left(\mathbf{u},v \right) =\int_{W}\int_T K^{\text{s}}_{\epsilon }\left(\mathbf{z}-\mathbf{u}\right) K^{\text{t}}_{\delta}\left(t-v\right) \de\mathbf{z} \de t.
	\end{equation} 
	
	The bandwidth selection for the estimation of spatio-temporal (or any dimension) kernels remains an open issue in statistics \citep{illian2008}, as there is no universal rule to set it; however, there are many methods to choose a bandwidth in the spatial case. Some of those methods rely on minimising error measures \citep{diggle1985kernel,berman1989estimating}, in rules of thumb \citep{scott2015multivariate,baddeley2015spatialR},  on likelihood cross validation criterion \citep{loader1999local}, or Campbell's formula \citep{cronie2018bandwidth}. In the specific spatio-temporal context, unfortunately, bandwidths have not been so profoundly covered.
	
	\subsubsection{Computation by convolution and FFT}\label{sec:FFTfixed:st}
	Let $\mathcal{N}$ a counting measure assigning 1 to every point of the point pattern $X=\{(\mathbf{u}_i,v_i)\}_{i=1}^n$, and let $K(\mathbf{u},v):=K^{\text{s}}_{\epsilon}\left(\mathbf{u}\right) K^{\text{t}}_{\delta}\left(v \right)$ for some fixed bandwidths $\epsilon$ and $\delta$ and $(\mathbf{u},v) \in W\times T$, then the convolution of $K$ with the counting measure $\mathcal{N}$ is given by
	\begin{eqnarray*}
		(K\ast \mathcal{N})(\mathbf{u},v)&=&\int_{\mathbb{R}^2 \times \mathbb{R}} K(\mathbf{u}-\mathbf{u'},v-v') \de \mathcal{N}(\mathbf{u'},v') \\
		&=&\sum_{i=1}^n K(\mathbf{u}-\mathbf{u}_i,v-v_i) \\
		&=& \tilde{\lambda}(\mathbf{u},v),
	\end{eqnarray*} 
	where $\tilde{\lambda}(\mathbf{u},v)$ is the numerator in Eq.\eqref{eq:stkernelintensity}.  For the edge correction, consider the indicator function $\mathbf{1}_{W\times T}(\mathbf{u},v)$ that assigns $1$ to every point of the region $W  \times T$ and $0$ elsewhere, then we consider the convolution of this function with $K$
	\begin{eqnarray*}
		(\mathbf{1}_{W\times T} \ast K)(\mathbf{u},v)&=&\int_{\mathbb{R}^2 \times \mathbb{R}} \mathbf{1}_{W\times T}(\mathbf{u'},v')  K(\mathbf{u}-\mathbf{u'},v-v') \de (\mathbf{u'},v') \\
		&=&\int_{W \times T} K(\mathbf{u}-\mathbf{u'},v-v') \de (\mathbf{u'},v') \\
		&=& e(\mathbf{u},v),
	\end{eqnarray*} 
	this idea goes back to \cite{koch2003spectral} and it was considered in the planar case by \cite{Davies2018kernel}. 
	
	A well-known numerical method used to compute the intensity fast and efficiently is first to discretise the data to a very fine grid, and then use the Fast Fourier transform to apply the convolution over the data with the kernel to obtain the estimate. This method is known as the \textit{binned estimator,} and \cite{silverman1982kernelfourier} first proposed it in the univariate context. Let $\mathcal{R}\supseteq W\times T$ be a three-dimensional rectangle enclosing $W\times T$; if we partition $\mathcal{R}$ into $M_1\times M_2 \times M_3$ rectangular voxels with equal volume $V=m_1 m_2 m_3$; let $\mathbf{c}_{ijk}$ be the centroid of the voxel $ijk$ ($i$th horizontal, $j$th vertical and $k$th temporal axes position). Let $I=\{1,\ldots,M_1\}$, $J=\{1,\ldots,M_2\}$ and $L= \{1,\ldots,M_3\}$ be the index sets; we define the grid $C=\{\mathbf{c}_{ijk}:i\in I, j\in J, k \in L \}$; there are several alternatives to discretise the counting measure $\mathcal{N}$ measure as a collection of weights $\omega_{ijk}$. The \textit{simple binning} involves assigning a unit mass to the grid point closest to every data point, i.e., $\omega_{ijk}$ is the number of observations of $X$ falling in the voxel $ijk$. On the other hand, the \textit{linear binning} \citep[see, e.g.,\ ][]{jonesandlotwick1983empiricalcharacteristic,wand1994kernelfast} assigns a fractional mass to the grid point depending on how far is the data point. In what follows, we use simple binning for simplicity. Once we have the weights calculated, an approximate form of the numerator of Eq. \eqref{eq:stkernelintensity} is 
	\begin{equation*}
		\tilde{\lambda}^* \left(\mathbf{c}_{ijk} \right) =\sum_{\ell_3=1}^{M_3} \sum_{\ell_2=1}^{M_2} \sum_{\ell_1=1}^{M_1}K\left(\mathbf{c}_{ijk}-\mathbf{c}_{\ell_1 \ell_2 \ell_3}\right)\omega_{\ell_1 \ell_2 \ell_3},
	\end{equation*}
	Note that $\omega_{\ell_1 \ell_2 \ell_3}=0$ outside of the set $I\times J \times L$, so after a straightforward change of variable we have that
	\begin{equation}\label{eq:lambdagrid}
		\tilde{\lambda}^* \left(\mathbf{c}_{ijk} \right) =\sum_{\ell_3=1-M3}^{M_3-1} \sum_{\ell_2=1-M_2}^{M_2-1} \sum_{\ell_1=1-M1}^{M_1-1}\omega_{i-\ell_1, j-\ell_2, k-\ell_3}K\left(m_1\ell_1, m_2 \ell_2, m_3 \ell_3\right),
	\end{equation}
	and for the edge correction factor we have that
	\begin{equation*}
		e^* \left(\mathbf{c}_{ijk} \right) =V\sum_{\ell_3=1-M3}^{M_3-1} \sum_{\ell_2=1-M_2}^{M_2-1} \sum_{\ell_1=1-M1}^{M_1-1}\mathbf{1}_{W\times T}\left(\mathbf{c}_{i-\ell_1, j-\ell_2, k-\ell_3}\right)K\left(m_1\ell_1, m_2 \ell_2, m_3 \ell_3\right).
	\end{equation*}
	A noteworthy point is that the array $\tilde{\lambda}^* \left(\mathbf{c}_{ijk} \right)$ and $e^* \left(\mathbf{c}_{ijk} \right) $ correspond to the discrete convolution of the $\omega$'s and the indicators with the kernel so that they can be computed rapidly using the fast Fourier transform (FFT). This algorithm only requires $O(M_1 M_2 M_3 \log(M_1) \log(M_2) \log(M_3) )$ operations outperforming the direct estimation that requires $O(M_1^2M_2^2M_3^2)$ operations \citep{wand1994kernelfast}.\\
	
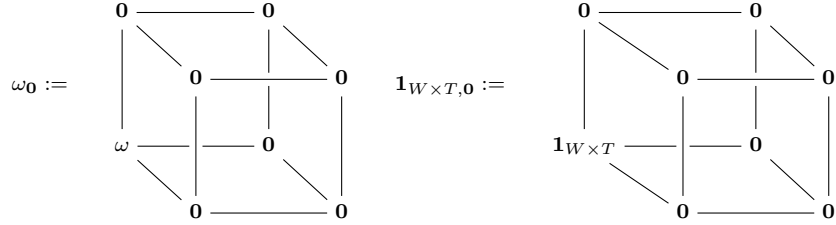
\begin{figure}[h]
		\centering
		\begin{tabular}{rlrl}
			$\omega_{\mathbf{0}}:=$ &
			\begin{tikzpicture}[baseline=0.3cm]
				\matrix[matrix of math nodes,column sep={2em},row sep={1.5em},
				text height=1.5ex,text depth=.5ex]
				{
					|(e)|  \mathbf{0} & & |(0)|   \mathbf{0}    \\
					& |(1)| \mathbf{0}    & & |(01)|  \mathbf{0}  \\
					|(2)|  \omega     & & |(02)|  \mathbf{0}  \\
					& |(12)| \mathbf{0}  & & |(012)| \mathbf{0} \\
				};
				
				\draw[-] (e)  to (2);
				\draw[-] (0)  to (02);
				\draw[-] (e)  to (0);
				\draw[-] (2)  to (02);
				
				\draw[-,over] (1)  to (12);
				\draw[-]      (01) to (012);
				\draw[-,over] (1)  to (01);
				\draw[-]      (12) to (012);
				
				\draw[-] (e)  to (1);
				\draw[-] (0)  to (01);
				\draw[-] (2)  to (12);
				\draw[-] (02) to (012);
			\end{tikzpicture}
			&$\mathbf{1}_{W\times T, \mathbf{0}}:=$ &
			\begin{tikzpicture}[baseline=0.3cm]
				\matrix[matrix of math nodes,column sep={2em},row sep={1.5em},
				text height=1.5ex,text depth=.5ex]
				{
					|(e)|  \mathbf{0} & & |(0)|   \mathbf{0}    \\
					& |(1)| \mathbf{0}    & & |(01)|  \mathbf{0}  \\
					|(2)|  \mathbf{1}_{W\times T}     & & |(02)|  \mathbf{0}  \\
					& |(12)| \mathbf{0}  & & |(012)| \mathbf{0} \\
				};
				
				\draw[-] (e)  to (2);
				\draw[-] (0)  to (02);
				\draw[-] (e)  to (0);
				\draw[-] (2)  to (02);
				
				\draw[-,over] (1)  to (12);
				\draw[-]      (01) to (012);
				\draw[-,over] (1)  to (01);
				\draw[-]      (12) to (012);
				
				\draw[-] (e)  to (1);
				\draw[-] (0)  to (01);
				\draw[-] (2)  to (12);
				\draw[-] (02) to (012);
			\end{tikzpicture}
		\end{tabular}
		\caption{\label{fig:cubew}Visual representation of the zero-padded weight matrix and indicator function, $\mathbf{0}$ represents a generic matrix of zeros.}
	\end{figure}

\begin{figure}[h]
		\begin{tabular}{p{0.04\linewidth} m{1.18\linewidth}}
			\small $K_{\mathbf{0}}:= \quad $
			& \includegraphics[width=0.8\linewidth]{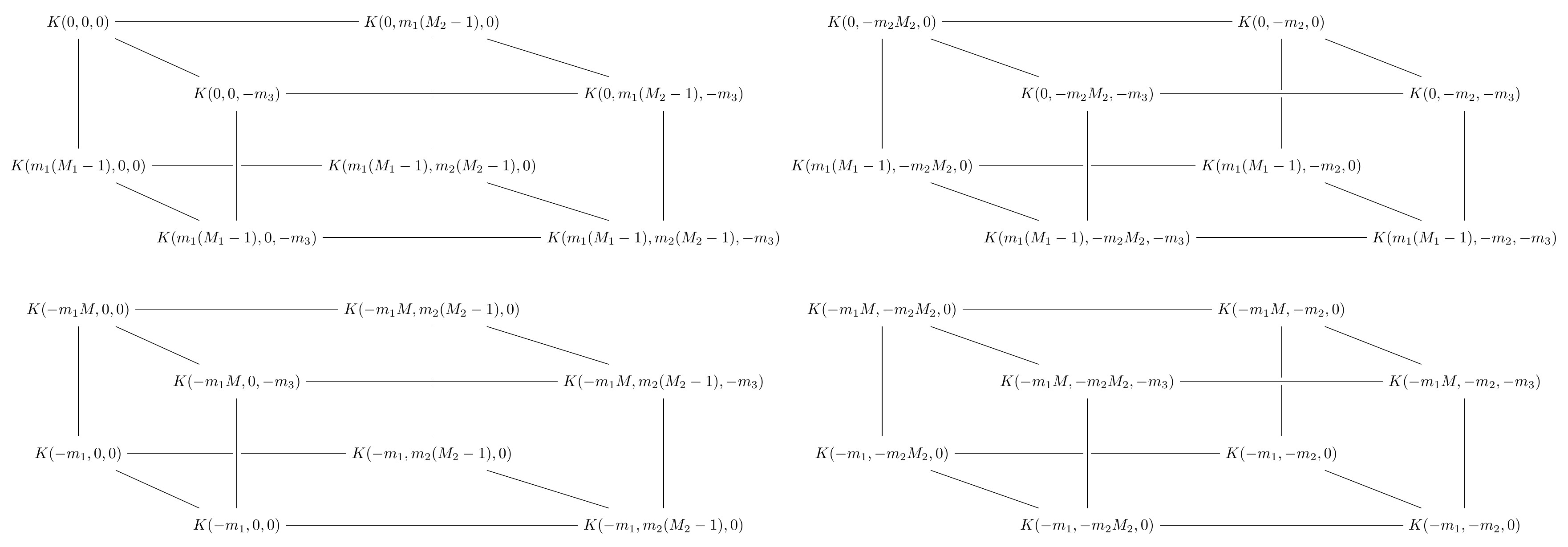}
		\end{tabular}
		\caption{\label{fig:K0} Two layers of the reflected kernel evaluation array at depth coordinates $0$ and $-m_3$.}
	\end{figure}

	Let $\mathscr{F}$ denote the discrete Fourier transform, $\mathscr{F}^{-1}$ its inverse and $f$ and $g$ two integrable functions, then, as a consequence of the convolution theorem $f\ast g = \mathscr{F}^{-1}(\mathscr{F}(f)\mathscr{F}(g))$. However, the theorem requires that the involved functions be periodic; we employ a zero-padding technique to ensure the periodicity. We consider larger arrays of sizes $P_{M_1}\times P_{M_3} \times P_{M_3}$ where $P_{M_i}=2M_i,i=1,2,3.$ The weights array $\omega_{\mathbf{0}}$ is an array of dimensions $P_{M_1}\times P_{M_3} \times P_{M_3}$ filled with zeros, except for the sub-array $\omega = (\omega_{ijk})_{i\in I, j\in J, k\in L}$ located in the lowest vertex. In the same way, we define a zero-padded array $\mathbf{1}_{ W\times  T, \mathbf{0}}$ for the edge correction; the schemes of these arrays are displayed in Fig. \ref{fig:cubew}. 
	We denote the reflected kernel evaluation array by $K_{\mathbf{0}}$; in this array, the kernel values are padded in a wrap-around order, and its dimensions are $P_{M_1} \times P_{M_2} \times P_{M_3}.$ Note that the dimensional index for the horizontal axis ranges from $0,\ldots, m_1(M_1-1), -m_1M_1,\ldots -m_1$, the vertical axis from $0,\ldots, m_2(M_2-1), -m_2M_2,\ldots -m_2$ and the depth axis from $0,\ldots, m_3(M_3-1), -m_3M_3,\ldots -m_3$. Fig. \ref{fig:K0} shows the padding scheme at two of the 2D layers of $K_{\mathbf{0}}$ at values $0$ and $-m_3$ (the first and the last).
	
	Finally, we gather all the ingredients to estimate the intensity by using the extended arrays Eq. \eqref{eq:lambdagrid} to obtain
	\begin{equation*}
		\tilde{\lambda}^* \left(\mathbf{c}_{ijk} \right) = \Re\left[\mathscr{F}^{-1}(\mathscr{F}(\omega_{\mathbf{0}})\mathscr{F}(K_{\mathbf{0}}))\right]_{ijk}; \quad i\in I, j\in J, k \in L,
	\end{equation*}
	and for the edge correction
	\begin{equation*}
		e^* \left(\mathbf{c}_{ijk} \right) = \Re\left[\mathscr{F}^{-1}(\mathscr{F}(\mathbf{1}_{W\times T, \mathbf{0}})\mathscr{F}(K_{\mathbf{0}}))\right]_{ijk}; \quad i\in I, j\in J, k \in L,
	\end{equation*}
	where $\Re[\cdot]$ represents the real component of a complex number required due to the numerical error \citep{Davies2018kernel}. By definition $\tilde{\lambda}^* \left(\mathbf{c}_{ijk} \right):=0, \forall \mathbf{c}_{ijk} \notin W\times T$ and the final spatio-temporal fixed estimator is given by 
	\begin{equation}\label{eq:FFTfixed}
		\hat{\lambda} \left(\mathbf{c}_{ijk} \right) =\frac{\tilde{\lambda}^* \left(\mathbf{c}_{ijk} \right) }{e^* \left(\mathbf{c}_{ijk} \right) }.
	\end{equation}
	
	\section{Spatio-temporal adaptive estimator}\label{sec:stadaptive}
	Although the adaptive estimator using kernels is well known in the multivariate context; i.e., when the coordinates to be smoothed have several dimensions, it has not been explicitly defined or studied in the spatio-temporal point processes framework. The Gaussian kernels make both the three-dimensional and the spatio-temporal estimators algebraically equivalent, although with different bandwidths \citep{wand1995kernel}. This case is usually ignored in the multivariate context or considered only in anisotropic settings. We define the spatio-temporal adaptive kernel estimator as follows,
	\begin{equation}
		\hat{\lambda}_{\epsilon, \delta}\left(\mathbf{u},v \right) =\frac{1}{e_{\epsilon,\delta} \left(\mathbf{u},v\right) }\sum_{i=1}^{n}{K^{\text{s}}_{\epsilon(\mathbf{u}_i)}\left(\mathbf{u}-\mathbf{u}_{i}\right)  K^{\text{t}}_{\delta(v_i)}\left(v-v_{i}\right) },\qquad (\mathbf{u}, v) \in W\times T,
		\label{eq:stadaptiveestimator}
	\end{equation}
	where 
	\begin{equation}
		e_{\epsilon,\delta} \left(\mathbf{u},v\right)  = \int_{W}\int_{T}K^{\text{s}}_{\epsilon(\mathbf{u'})}\left(\mathbf{u}-\mathbf{u'}\right)  K^{\text{t}}_{\delta(v')}\left(v-v'\right) \de \mathbf{u'} \de v',
	\end{equation}
	is the spatio-temporal edge correction inspired in the initially developed by \cite{marshalhazelton2010boundarieskernel}. This uniform edge correction attempts to fix the bias of the estimate at the edges of the study region (spatial and temporal), which can produce potentially unreliable estimations. Other edge correction alternatives, such as Diggle's edge correction for space-time (in which mass is preserved) \citep[see e.g.\ ][]{gonzalez2019tornadoes,ghorbani2021separability}, are compatible with our estimator and the partitioning technique presented in Section \ref{partitioning} to speed up the computation.  The bandwidth functions are defined as
	\begin{equation}\label{eq:abramsomband}
		\epsilon(\mathbf{u})=\frac{\epsilon^{\star}}{\gamma^{\text{s}}} \sqrt{\frac{n}{\lambda^{\text{s}}(\mathbf{u})}}, \quad \text{and} \quad
		\delta(v)=\frac{\delta^{\star}}{\gamma^{\text{t}}} \sqrt{\frac{n}{\lambda^{\text{t}}(v)}}, 
	\end{equation}
	where $\epsilon^{\star}$ and $\delta^{\star}$ stand for spatial and temporal smoothing multipliers known as \textit{global bandwidths}, $\lambda^{\text{s}}(\mathbf{u})$ and $\lambda^{\text{t}}(v)$ are marginal intensity functions in space and time, and $\gamma^{\text{s}}$ and $\gamma^{\text{t}}$ are the geometric mean terms for the marginal intensities evaluated in the points of the point pattern. This approach was proposed originally by \cite{abramson1982bandwidth} and requires pilot (fixed-bandwidth) estimates of the marginal intensities by using the global bandwidths; the inclusion of the geometric mean terms aims to free the bandwidth factor from the data scale, as pointed out by \cite{silverman1986density} and \cite{davieshazelton2010adaptiverisk}.
	
	\subsection{Fast estimation through a partitioning algorithm} \label{partitioning}
	To quickly calculate the adaptive estimator, our idea is to reduce the number of bandwidths, which in the direct estimator is $n$, to a much lower number, and therefore reduce the number of operations in that same proportion. For this purpose, we go to group the bandwidths in bins. Approximating an adaptive estimate through binning dates back to \cite{sainscott1996locallyadaptive} in the univariate case and to \cite{sain2002multivariateadaptive} in the multivariate case.  They used piecewise functions for the bandwidth, and they showed that their estimators are much better from a practical point of view as they offer good approximations and significantly reduce the number of operations while maintaining the mean integrated square error (MISE).

	To calculate the adaptive estimator given in Eq.\eqref{eq:stadaptiveestimator}, we follow the methodology proposed by \cite{Davies2018kernel}. This is based on the discretisation of the bandwidths chosen for each point through the empirical quartiles of its sampling distribution.  Starting from Eq. \eqref{eq:abramsomband}, we obtain a set of spatial bandwidths $\{\epsilon_1,\ldots,\epsilon_n\}$ and a set of temporal ones $\{\delta_1,\ldots,\delta_n\}$ by using pilot intensities calculated in the classical context. We consider the empirical $p$th quantiles, $\hat{\epsilon}^{(p)}$ and $\hat{\delta}^{(p)}$ of the bandwidths together with two \textit{quantile steps}, $\xi_1, \xi_2 \in (0,1],$ such that $C_1=\xi_1^{-1}$ and $C_2=\xi_2^{-1}$ are integers. We define the bandwidth bins through the values $\{\hat{\epsilon}^{(0)},\hat{\epsilon}^{(\xi_1)},\hat{\epsilon}^{(2 \xi_1)},\ldots,\hat{\epsilon}^{(1)}\}$ and $\{\hat{\delta}^{(0)},\hat{\delta}^{(\xi_2)},\hat{\delta}^{(2 \xi_2)},\ldots,\hat{\delta}^{(1)}\}$, and we place each observation $(\mathbf{u}_i,v_i)$ in one of the bins shown in Table \ref{tab:stbins}.
	\begin{table}[h!]
		\centering
		\resizebox{0.99 \linewidth}{!}{
			\opt{
				\begin{tabular}{  c  c  c  c }
					$\left[\hat{\epsilon}^{(0)} , \hat{\epsilon}^{(\xi_1)} \right] \times \left[\hat{\delta}^{(0)} , \hat{\delta}^{(\xi_2)} \right]$ & 
					$\left(\hat{\epsilon}^{(\xi_1)} , \hat{\epsilon}^{(2\xi_1)} \right] \times \left[\hat{\delta}^{(0)} , \hat{\delta}^{(\xi_2)} \right]$ & $\cdots$ &
					$\left(\hat{\epsilon}^{(\{C_1-1\}\xi_1)}, \hat{\epsilon}^{(1)}  \right] \times \left[\hat{\delta}^{(0)} , \hat{\delta}^{(\xi_2)} \right]$\\ [3mm]
					$\left[\hat{\epsilon}^{(0)} , \hat{\epsilon}^{(\xi_1)} \right] \times \left(\hat{\delta}^{(\xi_2)} , \hat{\delta}^{(2\xi_2)} \right]$& 
					$\left(\hat{\epsilon}^{(\xi_1)} , \hat{\epsilon}^{(2\xi_1)} \right] \times \left(\hat{\delta}^{(\xi_2)} , \hat{\delta}^{(2\xi_2)} \right]$ & $\cdots$ &
					$\left(\hat{\epsilon}^{(\{C_1-1\}\xi_1)}, \hat{\epsilon}^{(1)}  \right]  \times \left(\hat{\delta}^{(\xi_2)} , \hat{\delta}^{(2\xi_2)} \right]$ \\
					$\vdots$ & $\vdots$ & $\ddots$ & $\vdots$\\ 
					$\left[\hat{\epsilon}^{(0)} , \hat{\epsilon}^{(\xi_1)} \right] \times \left(\hat{\delta}^{(\{C_2-1\})\xi_2} , \hat{\delta}^{(1)} \right]$&
					$\left(\hat{\epsilon}^{(\xi_1)} , \hat{\epsilon}^{(2\xi_1)} \right] \times \left(\hat{\delta}^{(\{C_2-1\})\xi_2} , \hat{\delta}^{(1)} \right]$&$\cdots$&
					$\left(\hat{\epsilon}^{(\{C_1-1\}\xi_1)}, \hat{\epsilon}^{(1)}  \right]  \times \left(\hat{\delta}^{(\{C_2-1\})\xi_2} , \hat{\delta}^{(1)} \right]$
				\end{tabular}
			}
		}
		\caption{\label{tab:stbins} Array of spatio-temporal discretised bandwidth bins.}
	\end{table}
	
	The sets of bins generate a disjoint partition of the original point pattern $X$ into $C_1\times C_2$ sets $Y_{ij}$ and 
	$$X=\bigcup_{ij} Y_{ij}.$$
	Suppose that in each subset $Y_{ij},$ the intensity is estimated using a bandwidth defined as the midpoint of the respective bin where it belongs. In that case, the intensity can be approximated as
	\begin{equation}\label{eq:partitioninglambda}
		\hat{\lambda}_{\epsilon,\delta}\left(\mathbf{u}, v \right) \approx \sum_{i =1}^{C_1} \sum_{j=1}^{C_2} \hat{\lambda}^{*}_{\bar{\epsilon}_i,\bar{\delta}_j}\left(\mathbf{u},v | Y_{ij} \right),
	\end{equation}
	where $\bar{\epsilon}_i$ and $\bar{\delta}_j$ represent the midpoints of the $i$th spatial and $j$th temporal bins and $\hat{\lambda}^{*}_{\bar{\epsilon}_i,\bar{\delta}_j}\left(\mathbf{u},v | Y_{ij} \right)$ is a fixed-bandwidth estimate based on the sub-pattern $Y_{ij}$. Each of the addends on the right-hand side of Eq. \eqref{eq:partitioninglambda} is calculated by the procedure described in Section \ref{sec:FFTfixed:st}, that is, by Fourier transforms and Eq. \eqref{eq:FFTfixed}. Notice that in each addend we include the corresponding edge correction factor.
	
	\section{Simulation study}\label{sec:simulation}
	In this section, we perform a simulation study to assess the performance of the partition method described in Section \ref{sec:stadaptive} compared with the direct method, which consists of evaluating the kernel at each data point and each point of the grid. For a single simulation, we generate a spatio-temporal point pattern as a realisation of a spatio-temporal log-Gaussian Cox process. Then we estimate the spatial and temporal global bandwidths and compute the pilot marginal intensities using the projection of point pattern coordinates to space and time, respectively. Finally, we compute the adaptive intensity through the direct and partitioning algorithms, the latter for different binning schemes. Along with each estimation, we calculate the wall timing clock.
	
	We start generating the point patterns, which are observations of point processes with stochastic intensities. Let $\{Z(\mathbf{u}, v)\}$ be a real-valued Gaussian random field with mean $\mu(\mathbf{u},v)$ and covariance $\Cov(Z(\mathbf{u}_i,v_i),Z(\mathbf{u}_j,v_j))$, and let $\Lambda(\mathbf{u},v)=\exp{(Z(\mathbf{u},v))}$ be the intensity. Then, a log-Gaussian Cox process is a Poisson point process in which the intensity function is a realisation of $\Lambda(\mathbf{u},v)$, say $\lambda(\mathbf{u},v)$ \citep[see e.g.,\ ][]{moller1998loggaussian,gabriel2013stpp,diggle2013book,diggle2013paradigm,gonzalezetal2016}. In this case, we employ an non-separable stationary covariance function of the form \citep{gneiting2006covariances},
	\begin{equation}\label{eq:geneitingcovariance}
		\Cov(\mathbf{u},v)=\frac{1}{\psi^2(v^2)}\phi \left(\frac{||\mathbf{u}||^2}{\psi(v^2)}\right), \quad \mathbf{u}\in\mathbb{R}^2, v \in \mathbb{R},
	\end{equation}
	where $\phi(r), r>0$, is a strictly monotone function, and $\psi(v),v>0$ is a positive function with a completely monotone derivative. In this case we use the so-called \textit{stable model}, i.e., $\phi(r)=\sigma^2 \exp(-cr)$ where $c$ is a non-negative scale parameter for space and $\sigma^2$ is the spatio-temporal variance, and 
	\begin{equation*}
		\psi(v)=\left(1 + a v^{\alpha}\right)^{\beta}, \quad \alpha,\beta \in (0,1],
	\end{equation*}
	were $a$ is non-negative scale parameter for time, $\alpha$ is an smoothness parameter and $\beta$ is the space-time interaction parameter. We fix three sets of parameters for the simulation that encompass some first-order non-separable point processes with different clustering degrees;
	\begin{enumerate}
		\item[Set 1.] $\sigma^2 = 2, c = 0.5, a = 0.8$ and $\alpha = \beta = 0.1$
		\item[Set 2.] $\sigma^2 = 0.5, c = 0.5, a = 0.8$ and $\alpha = \beta = 0.1$
		\item[Set 3.] $\sigma^2 = 0.5, c = 0.5, a = 0.8$ and $\alpha = \beta = 0.8$
	\end{enumerate}
	
\begin{figure}[h!tb]
		\includegraphics[width=\linewidth]{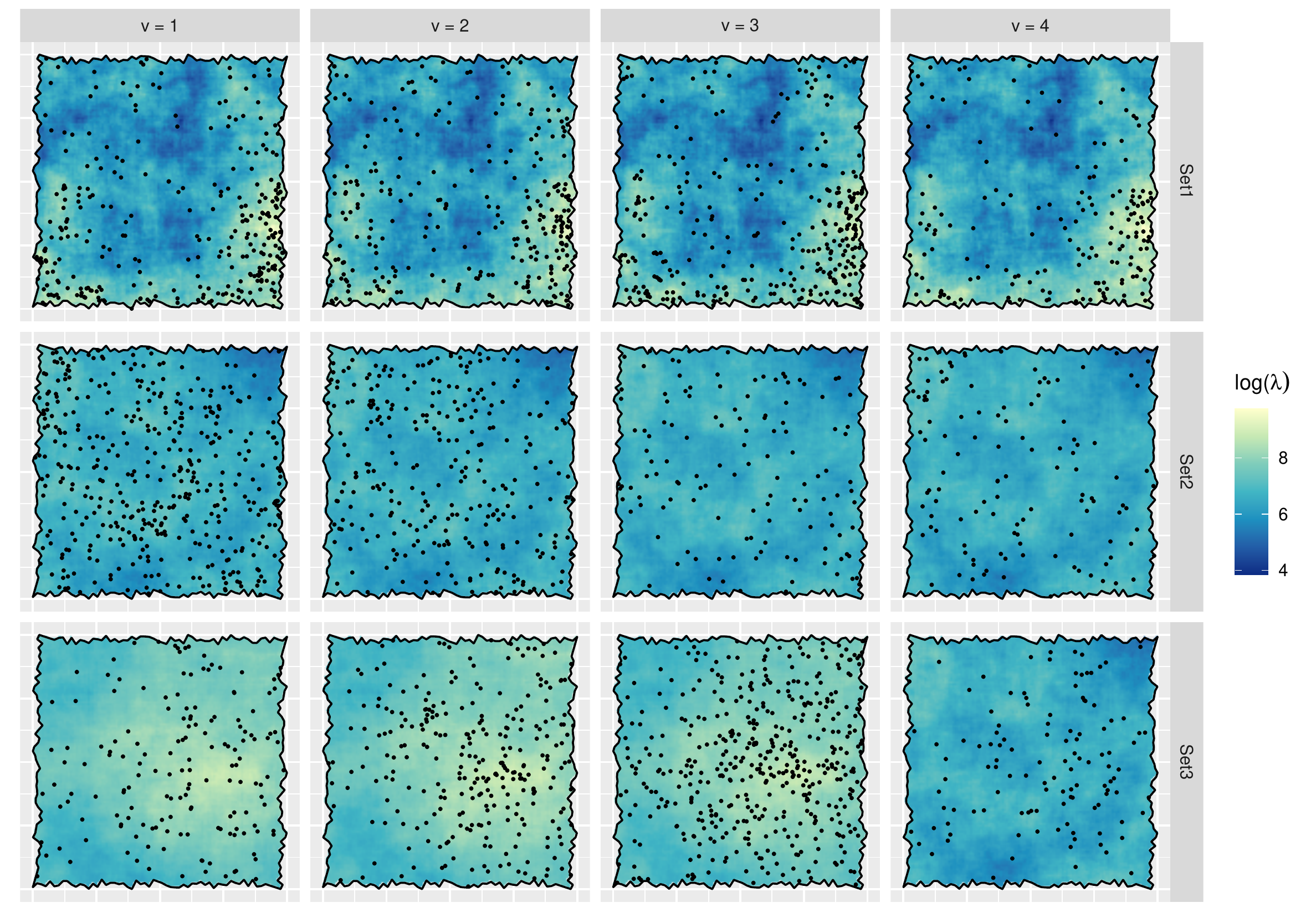}
		\caption{Spatial snapshots of three realisation of a spatio-temporal log-Gaussian point pattern with their underlying intensity function having non-separable Gneiting covariance functions. The irregular spatial window is enclosed in the unit square.}
		\label{fig:simulated_field}
	\end{figure}

	Figure \ref{fig:simulated_field} displays three realisations of spatio-temporal log-Gaussian Cox processes coming from Set 1, 2 and 3, with their underlined intensity functions, where the numbers of points come from a Poisson distribution with a mean of 1000.
	We generate 1000 realisations of log-Gaussian Cox processes with roughly 1000 points in a region $W\times T$, where $W$ is an irregular polygon contained in the unit square, and $T$ is the unit interval; the covariance of the Gaussian random field is Gneiting type (see, Eq. \eqref{eq:geneitingcovariance}). 
	
	The next step in the simulation process is estimating the global spatial bandwidths; we use the maximal smoothing principle \citep{terrell1990maximal,davies2018relativerisksparr} that consists of upper bounding the optimal bandwidths with respect to the asymptotic mean integrate squared error (MISE) of the density estimates to obtain oversmoothing bandwidths. These bandwidths are given by
	$
	\epsilon^{\star}\approx 1.085 \hat{\sigma} n^{-1/6},
	$
	where $\hat{\sigma}$ is the minimum half sum between the axis-specific standard deviations and the interquartile ranges normalised by 1.34 \citep{silverman1986density}. For the global temporal bandwidths, we use the method of \cite{sheather1991bandwidth}, which relies on the estimation of the density derivatives.
	
	Now, for computing the adaptive smoothing bandwidth functions (Eq. \eqref{eq:abramsomband}), we employ two separated pilot estimates, one for space and one for time computed using convolutions and FFT in two and one dimensions, respectively. 
	
	Finally, for every spatio-temporal point pattern, we estimate the intensity by using the partitioning algorithm by combining the number of bins in space and time, i.e., using all combinations of $\delta,\beta \in \{0.1, 0.05, 0.025, 0.01\}$. We use a spatio-temporal resolution of $128\times 128 \times 64$; the spatial resolution is a default in the \texttt{spatstat} and \texttt{sparr} packages \citep{baddeley2015spatialR,davies2018relativerisksparr}, and the temporal resolution was chosen as a compromise between the spatial axes.

	As we wish to compare the performance of our estimator (by assigning different amounts of spatio-temporal bins) we opt for the $L_2$-norm, which in the densities context is known as \textit{integrated squared error} (ISE), and it is defined as \citep{scott2015multivariate}
	\begin{equation*}
		\text{ISE}[\hat{f}]:= \int_W \int_{T} \left(\hat{f}(\mathbf{u},v) - f(\mathbf{u},v) \right)^2 \de \mathbf{u} \de v,
	\end{equation*}
	where $\hat{f}$ is the estimate using the partitioning algorithm and $f$ is the direct estimate, both of them expressed in on density scale, i.e., normalising so that they integrate one. 
	
\begin{figure}[h!]
		\includegraphics[width=\linewidth]{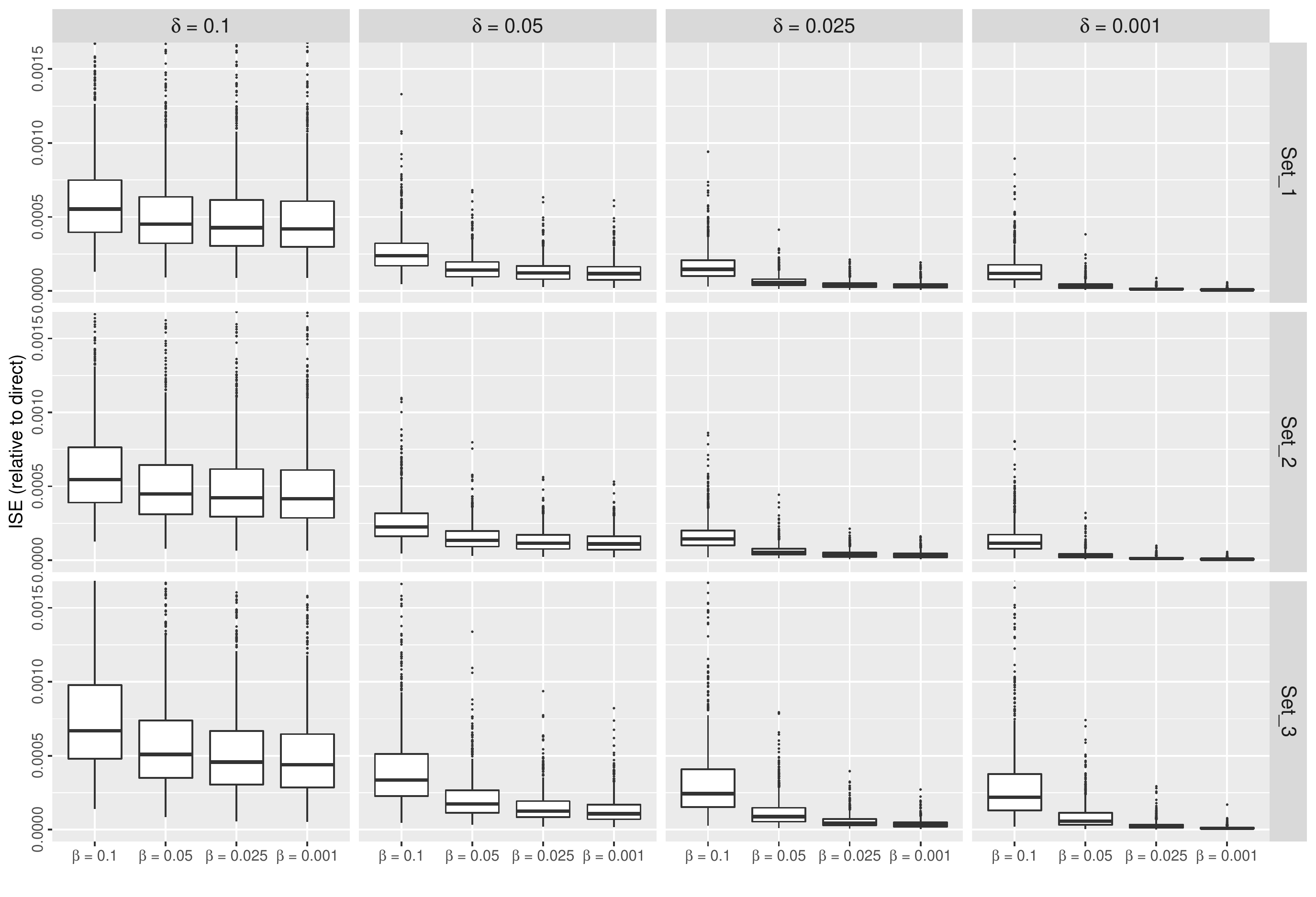}
		\caption{Integrated square errors with respect to direct estimation of intensity function using the partitioning algortithm. We consider combinations of the spatial $(\delta)$ and temporal $(\beta)$ bins ranging from 10 to 100 bins. The mean number of points in each simulated point pattern is $\mu = 1000$ and the resolution of the intensity arrays is $128\times 128\times 64$. We truncate the most extremes outliers for better understanding.}
		\label{fig:binsISE}
	\end{figure}

	Figure \ref{fig:binsISE} shows the ISE boxplots for each collection of estimates across bandwidth partitions. As expected and in line with \cite{Davies2018kernel}, coarser space-time partitions lead to larger errors. However, it is worth mentioning that even the coarsest partition, i.e., ten spatial bins and ten temporal bins $(\delta = 0.1, \beta = 0.1)$, generates an excellent estimate in terms of error. As expected, the estimates get closer to the direct estimate as the partitions get finer. 
	
\begin{figure}[h!]
		\includegraphics[width=\linewidth]{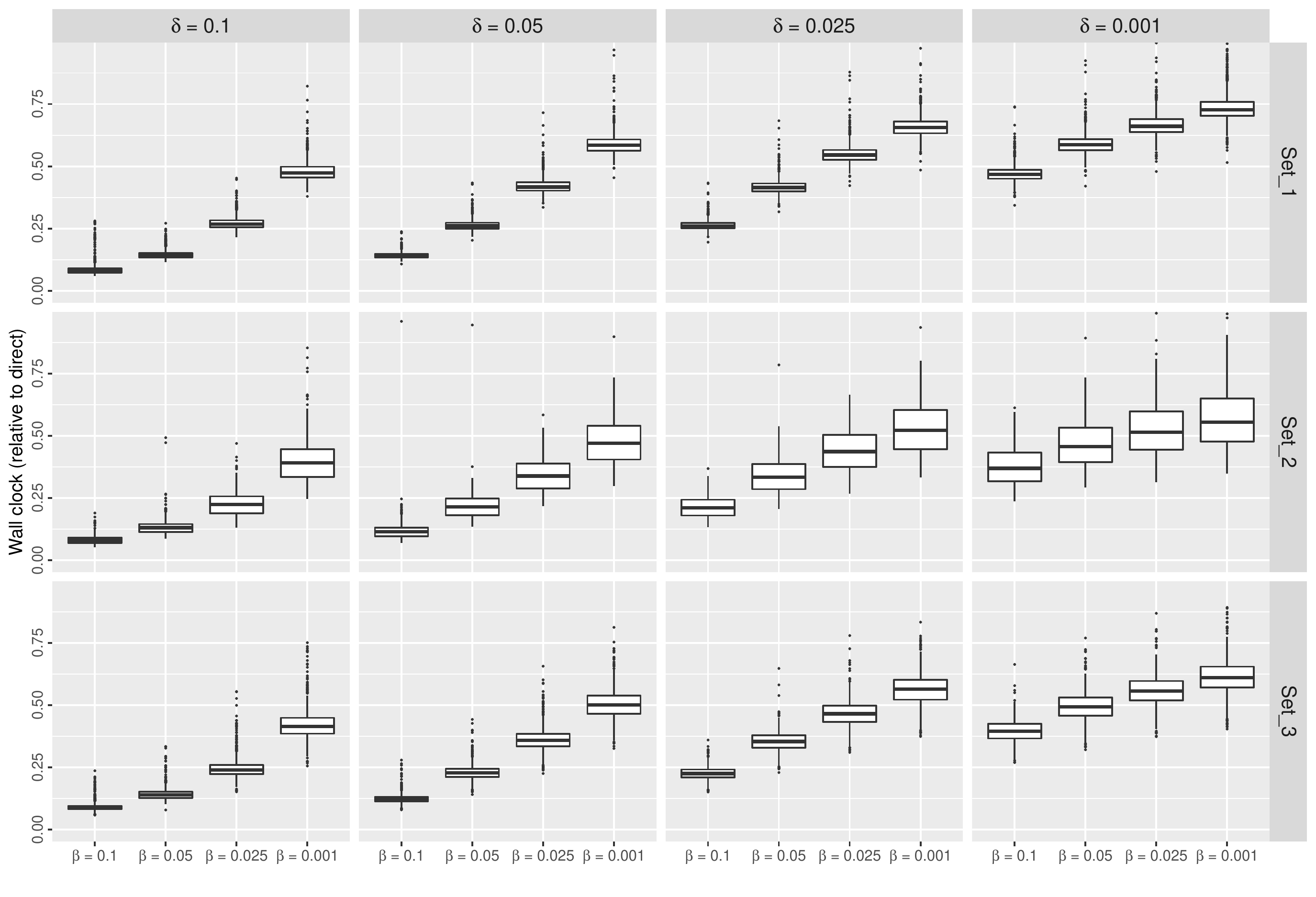}
		\caption{Wall clock timings relative to direct estimation using the partitioning algortithm.}
		\label{fig:binsclock}
	\end{figure}

	We also observe how elapsed execution times and bins increase together (Figure \ref{fig:binsclock}). However, in general terms, these times are always minimal compared to the time taken by direct estimation, implying that the binning procedure is always a considerably better solution in terms of execution times. It should be noted that when the number of spatial and temporal bins is maximum $(\delta=0.01,\beta=0.01)$, the time is considerably longer, and the gains in terms of error do not seem to be justified.
	
	\section{Application: Amazonia fires}\label{sec:amazonia}
	All around the globe, fires can affect all biomes significantly producing severe alterations to the primary ecosystems \citep{andela2019global}.  Fire activity in Amazonia varies considerably from month to month, likely influenced by human activity and climate change. The deployment of preventive and corrective measures can be aided by adequately describing the spatio-temporal evolution of the number of fires in every region of the Amazonia biome. Here we estimate the intensity function of the point pattern of wildfire ignition points observed in the Amazonian biome. The area covers almost the complete Amazon basin and contains the Amazon rainforest, a part of the tropical rainforest,  and some adjacent areas to the north and east. 
	
	The Amazon Dashboard (\url{http://www.globalfiredata.org}) tracks individual fires in the Amazon region using an cluster approach and classify active fire detections by fire type. Known as VIIRS, The Visible Infrared Imaging Radiometer Suite 375 m thermal anomalies provides data from a sensor aboard the joint NASA/NOAA Suomi National Polar-orbiting Partnership (Suomi NPP) and NOAA-20 satellites. The data are updated on a daily basis and there are several characteristics recorded with each single fire. For this analysis we select new and active deforestation fires (starting within past 24 hours) and have a sample of $59,910$ fires from 01/01/2021 to 10/10/2021 (284 days).
	
\begin{figure}[h!]
		\centering
		\includegraphics[width=0.75\linewidth]{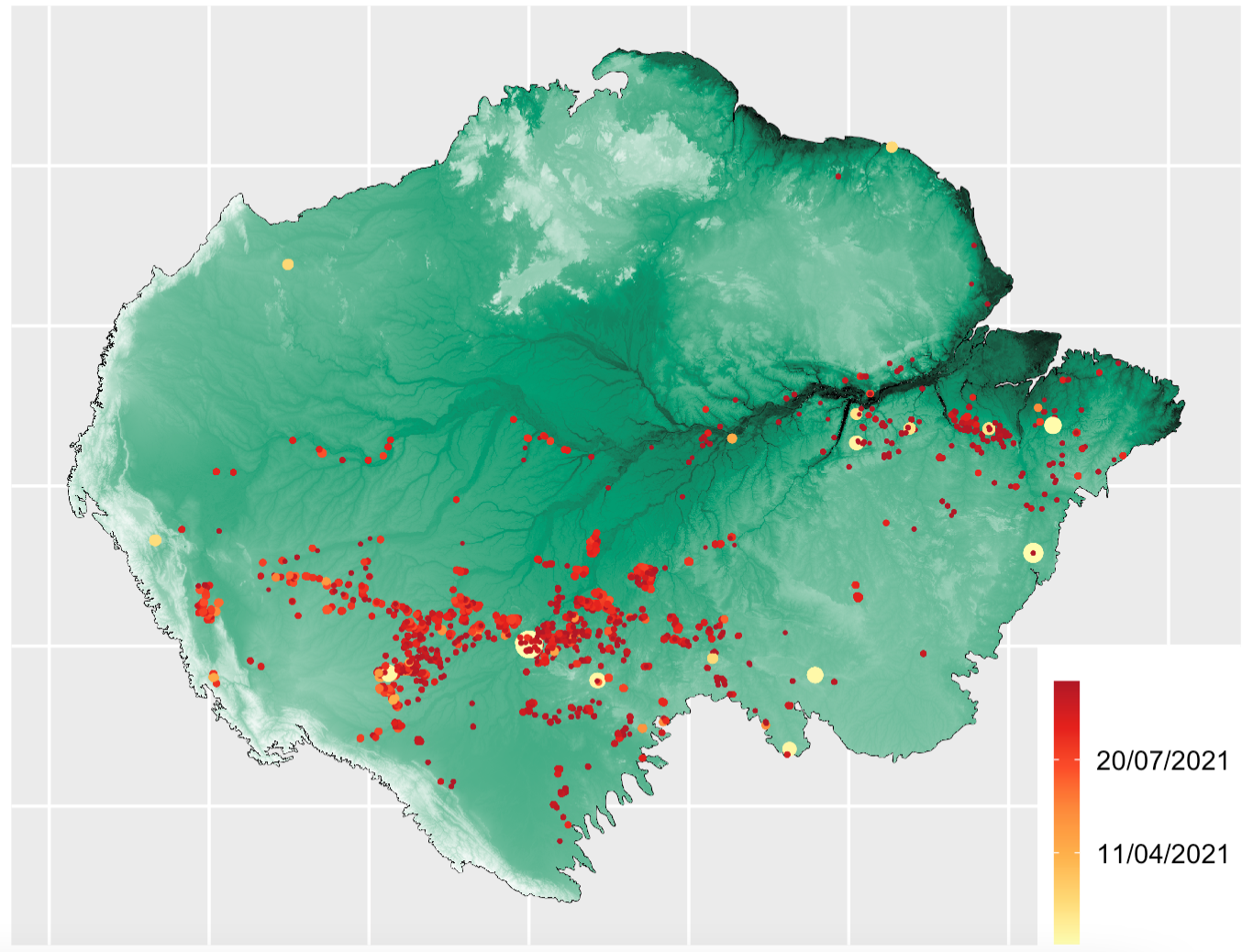}\\
		\includegraphics[width=0.75\linewidth]{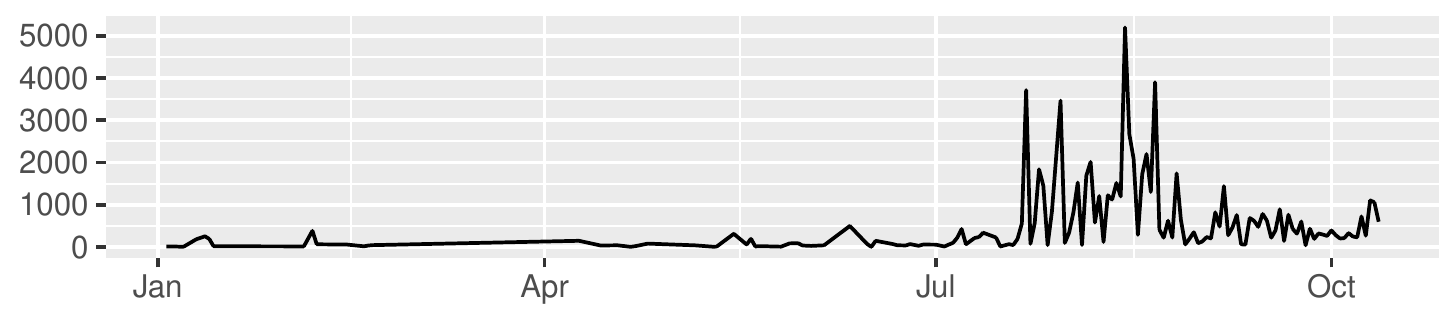}\\
		\caption{Deforestation fires in the Amazonia biome from 01/01/2021 to 10/10/2021, wider lighter points corresponds to older fires, and the background corresponds to the elevation of the Amazonia biome. The bottom panel shows the time series of fire counts throughout the sampling days.}
		\label{fig:amazonelevation}
	\end{figure}

	Figure \ref{fig:amazonelevation}, top panel, displays the locations of fires in the Amazon.
	Although it seems that there are few points in the map, this is a visual delusion due to two reasons. The first is that the  Amazon area is so large that it is difficult to appreciate what happens on such a small scale; the second is the high concentrations of points in minimal areas. Many fires seem to occur near each other, resulting in many places showing no fire activity at all while all of this activity is concentrated in a few locations. The pattern of temporal counts (Figure \ref{fig:amazonelevation}, top panel) allows us to appreciate that temporally, the pattern also lacks uniformity, being the month of August particularly prone to fires with much more pronounced variations than in any other month. This high inhomogeneity in the fire pattern makes estimating the intensity very difficult if a fixed bandwidth was chosen, which is why we resorted to the variable bandwidth kernel estimator discussed in this work to carry out the estimation.
	
	On the other hand, given that the initial coordinates of both the fires and the map are given in longitude-latitude, we have decided to use the Albers equivalent conic projection, which preserves the areas, to estimate the intensity.
	
	We first used the partitioning algorithm described in Section \ref{partitioning}. We chose 38 groups for the spatial bandwidths $(n^{1/3})$ and 6 for the temporal ones $(n^{1/6})$; we select the numbers of groups this way to ensure that the total number of groups is not bigger than $\sqrt{n}\approx 245$ meaning that so that the computation time is only about 228 times slower than fixed-bandwidth smoothing but roughly 59682 $(n-\sqrt{n})$ times faster that the direct estimation. The estimation took less than three minutes on a conventional computer. 
	
\begin{figure}[h!]
		\includegraphics[width=\linewidth]{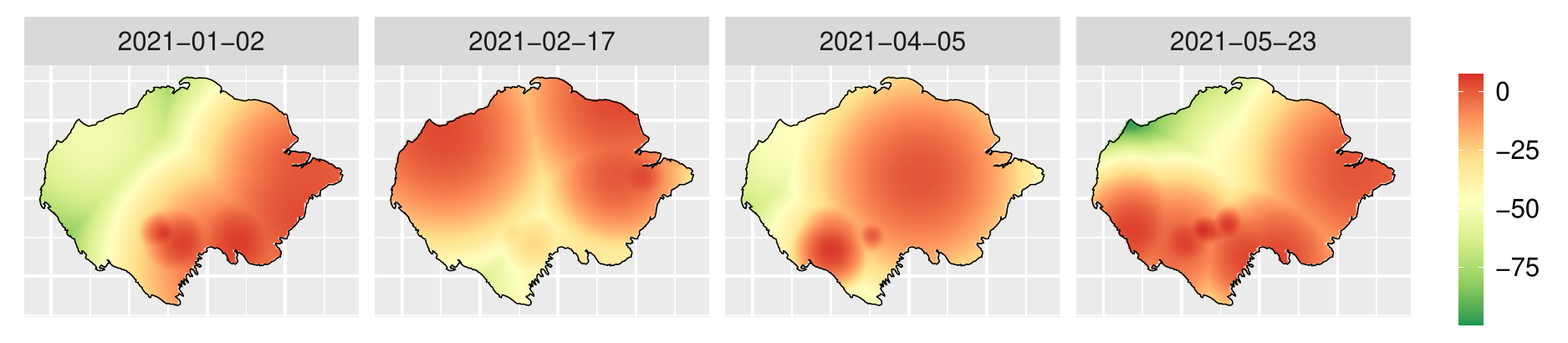}
		\includegraphics[width=\linewidth]{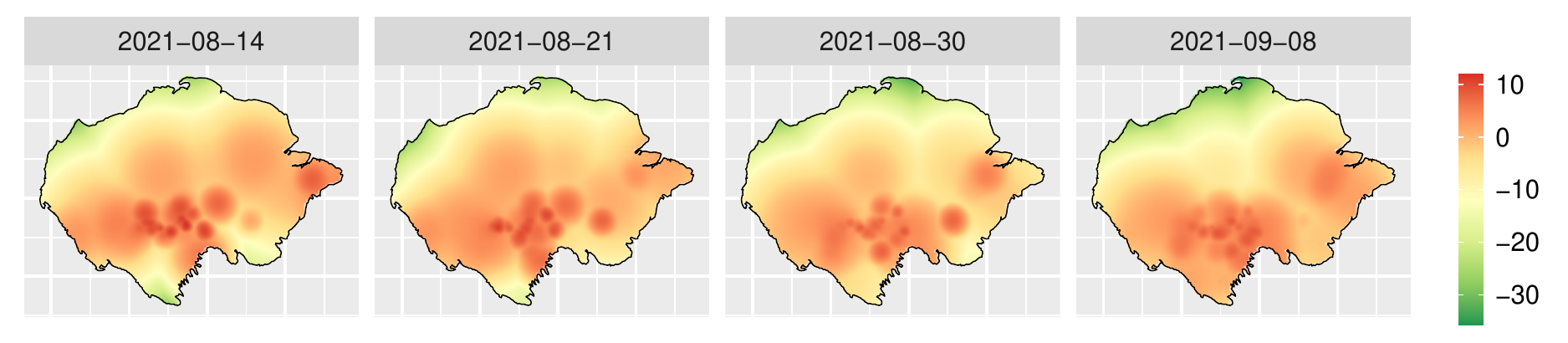}
		\caption{Snapshots of the intensity fires (in a log scale) in Amazon region upon eight specific times during the study period during . This estimation was carried out through the partitioning algorithm described in Section \ref{partitioning}. The upper row corresponds to the first semester of the year and the lower one to the second.}
		\label{fig:amazonintensity}
	\end{figure}

	A sample with some snapshots of the estimated spatio-temporal intensity can be seen in Figure \ref{fig:amazonintensity}. The two panels of the figure have different scales since they correspond to periods with hugely different densities in magnitude; if we chose a single scale, the drawings would probably flatten out and be visually indistinguishable.
	As expected, the local changes in the number of fires per area from one moment to another show that the intensity is not separable. We can also observe that, although the majority of the fires are concentrated in the south-central part, that is, part of Brazil, Bolivia, Paraguay and Peru; the south-eastern part also has large concentrations of fires; the northwestern part is the least prone, and this low number extends throughout the north of the biome. The fires are exacerbated in number during the warmer months, reaching spots with entire kilometres burning.

	\section{Discussion}\label{sec:discussion}
	In this paper, we have presented an adaptive estimator for the first-order intensity of a spatio-temporal point process; this estimator is non-parametric and is based on only one realisation of the point process to be computed. Given the nature of the estimator, in which every point is equipped with a particular bandwidth, a possible direct estimation is computationally unfeasible, and we have proposed an algorithm to perform this estimation in a fast way. This algorithm is an extension of the one presented by \cite{Davies2018kernel} for the spatial case, and it bases on the discretisation of the candidate bandwidths for each point in a point pattern. We have proven via simulation that our algorithm performs well in integrated squared error and computation time. Finally, we have applied our method to estimate the intensity of deforestation fires in the Amazon.

	An alternative computational approach could be reformulating the adaptive smoothing for a point pattern in $\mathbb{R}^2\times \mathbb{R}$ into one convolution in $\mathbb{R}^5$, and calculating it using the FFT approach. \cite{Davies2018kernel} conceived the adaptive estimation problem in the planar case $(\mathbb{R}^2)$ to a convolution problem in three-dimensional scale-space $(\mathbb{R}^3)$ \citep{chaudhuri2000scalespace}. This method dramatically reduced computational costs when performing the adaptive estimation. Unfortunately, when we talk about the spatio-temporal case, things work well in theory but not so well in practice. The key idea would be the following. We use the scale-space concept and define a five-dimensional kernel in scale-space as
	\begin{equation*}
		\mathscr{K}(\mathbf{u},\epsilon,t,\delta)=K^2_{\exp(-\epsilon)}(\mathbf{u})K^1_{\exp(-\delta)}(t),
	\end{equation*}
	where $K^2$ and $K^1$ represent the two- and one-dimensional Gaussian kernel functions \citep{silverman1986density}. Consider once again a counting measure $\mathcal{N}$ that assigns 1 to each of the points $\{\mathbf{u}_i,\log{(\epsilon_i)},t_i,\log{(\delta_i)} \}_{i=1}^n$, and then, consider the convolution of the kernel with the measure,
	\begin{eqnarray*}
		(\mathscr{K}\ast \mathcal{N})(\mathbf{u},\epsilon,t,\delta)&=&\int_{\mathbb{R}^5}\mathscr{K}(\mathbf{u}-\mathbf{u}_0,\epsilon - \epsilon_0,t-t_0,\delta-\delta_0) \de \mathcal{N}(\mathbf{u}_0,\epsilon_0,t_0,\delta_0) \\
		&=&\sum_{i=1}^n \mathscr{K}(\mathbf{u}-\mathbf{u}_i,\epsilon - \log(\epsilon_i),t-t_i,\delta-\log(\delta'_i)) \\
		&=& \sum_{i=1}^n  K^2_{\epsilon_i \exp(-\epsilon)}(\mathbf{u}-\mathbf{u}_i)K^1_{\delta_i \exp(-\delta)}(t-t'_i),
	\end{eqnarray*} 
	and evaluating the convolution in the hyperplane $\epsilon = \delta = 0$, we obtain
	$$(\mathscr{K}\ast \mathcal{N})(\mathbf{u},0,t,0)=\hat{\lambda}^*_{\epsilon, \delta}(\mathbf{u},t).$$
	Analogously, the edge-correction term may be expressed as a convolution sliced at the corresponding hyperplane.
	The previous expressions constitute five-dimensional functions, complicating things in terms of computational resources. For example, the padding step to make the function periodic must add  $M_1 M_2 M_3 M_4 M_5$ zeros to the kernel array and the counting measure array. Consequently, the memory expense and the number of operations to calculate the transform, its inverse, and the kernel's product with the counting measure increase exponentially. Thus, it causes a conventional computer to exceed its memory limits quickly, and therefore, the algorithm is useless in practice.
	\cite{scott2015multivariate} indicated this problem in several dimensions and recommended a practical limit of three dimensions; the advice is not to go beyond the planar case developed by \cite{Davies2018kernel}.
	
	One of the key points for the adaptive estimation of the intensity is the selection of the pilot intensity and the global bandwidth to use Abramson's rule. In this work, we have chosen spatial and temporal pilot intensities separated and estimated using the traditional method with fixed bandwidth, that is, the method through Fourier transforms in $\mathbb{R}$ for the temporal case and in $\mathbb{R}^2$ for the spatial case. However, using a non-separable estimator such as Eq. \ref{eq:FFTfixed}, could also be considered , and then integrating to obtain the marginal intensities. The method is not the fastest computationally speaking but could eventually improve the bandwidths for each data point.
	
	Since the edge correction is a much smoother surface than the intensity itself as in most regions it usually takes the value of one; \cite{Davies2018kernel} estimate this edge correction using a coarser mesh than the one used for the numerator of the intensity. They then use simple interpolation techniques to scale the surface to the intensity scale; this idea stems from the need for faster estimations. Although we have used the same meshes for both the intensity and the edge correction factor in the present work, exploring this possibility of coarser evaluation grids for the edge correction in space-time could be interesting.
	
	To conclude, the adaptive spatio-temporal estimator presented here allows us to face the first-order intensity estimation problem, especially in point patterns with high concentrations of points in certain areas, that is, those highly heterogeneous, providing each point with its bandwidth. Although the complexity of the estimator is high, we can alleviate the computational burden through the partitioning algorithm we have introduced. In addition, we also provide the \texttt{R} code to easily and fast analyse point patterns that can arise in many scientific research fields.

	\bibliographystyle{imsart-nameyear} 
	\bibliography{bibliography}       
	
	
\end{document}